%% file: main.tex
\documentclass{article}

\usepackage[utf8]{inputenc}
\usepackage[T1]{fontenc}
\usepackage{microtype}
\usepackage{graphicx}
\graphicspath{{figures/}}
\usepackage{subcaption}
\usepackage{booktabs}
\usepackage{hyperref}
\usepackage{url}

\usepackage[accepted]{icml2026}

\usepackage{amsmath}
\usepackage{amssymb}
\usepackage{mathtools}
\usepackage{amsthm}
\usepackage[capitalize,noabbrev]{cleveref}

\usepackage{array}
\usepackage{tabularx}
\usepackage{multirow}
\usepackage{float}
\usepackage{enumitem}
\usepackage{alphalph}

\newcommand{\meg}{\mathrm{MEG}}
\newcommand{\mig}{\mathrm{MIG}}

\newcommand{\authornoteA}[1]{}
\newcommand{\authornoteB}[1]{}
\newcommand{\authornoteC}[1]{}
\newcommand{\para}[1]{\noindent \textbf{#1}}

\icmltitlerunning{The Interaction Tax: When Communication Erases Diversity in Multi-Agent Teams}

\begin{document}

\makeatletter
\renewcommand{\@pa}[1]{%
  \ifcsname the@affil#1\endcsname\else
    \ifcsname @icmlsymbol#1\endcsname\else
      \stepcounter{@affiliationcounter}%
      \newcounter{@affil#1}%
      \setcounter{@affil#1}{\value{@affiliationcounter}}%
    \fi
  \fi
}
\makeatother

\twocolumn[
  \icmltitle{The Interaction Tax: When Communication Erases Diversity \\in Multi-Agent Teams}

  \icmlsetsymbol{equal}{*}

  \begin{icmlauthorlist}
    \icmlauthor{Summer Eunhyung Ann}{uchi}
    \icmlauthor{Haokun Liu}{uchi}
    \icmlauthor{Chenhao Tan}{uchi}
  \end{icmlauthorlist}

  \icmlaffiliation{uchi}{Department of Computer Science, University of Chicago, Chicago, IL, USA}

  \icmlcorrespondingauthor{Summer Eunhyung Ann}{summereunann@uchicago.edu}

  \icmlkeywords{multi-agent LLMs, AI-assisted research workflows, scientific optimization,
    hidden evaluation, diversity, ensemble learning}

  \vskip 0.3in
]

\newcommand{\haokun}[1]{}
\newcommand{\summer}[1]{}
\newcommand{\chenhao}[1]{\textcolor{blue}{#1}}

\printAffiliationsAndNotice{}

\begin{abstract}
Does multi-agent LLM interaction help or hurt?  Some work reports gains from debate \citep{du2024debate}, critique loops \citep{chen2025magicore}, and mixture-of-agents synthesis \citep{wang2025moa}, while other work finds that interaction adds cost without improving quality under equal budgets \citep{tran2026equal,xu2026single,jarrett2025multiagent}, or that independent sampling already captures multi-agent gains \citep{li2024moreagents}. We argue this contradiction reflects a missing distinction, because not all multi-agent communication is equal. Different model families find structurally different solutions, but when agents read each other's complete outputs, their proposals converge within one round, erasing the diversity that motivates using multiple models. We call this the \emph{interaction tax}. We test 11 verifier-scored optimization tasks under matched budgets and find that full-solution interaction is a weak default. Independent proposal generation avoids this collapse. Full-solution interaction mainly makes agents stay close to the first solution they see instead of trying different approaches, and critique helps only if the violated rule is easy for the LLM to find and fix.
These results suggest that multi-agent performance depends less on the number of agents than on the information they exchange, and interaction helps only when agents share the right information at the right time.

\end{abstract}

\begin{figure*}[t]
\centering
\begin{subfigure}{0.31\textwidth}
\includegraphics[width=\textwidth]{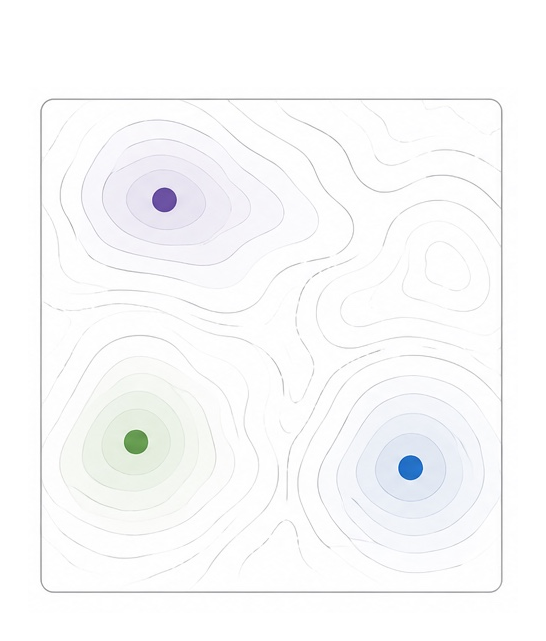}
\caption{Independent proposals}
\end{subfigure}
\hfill
\begin{subfigure}{0.34\textwidth}
\includegraphics[width=0.91\textwidth]{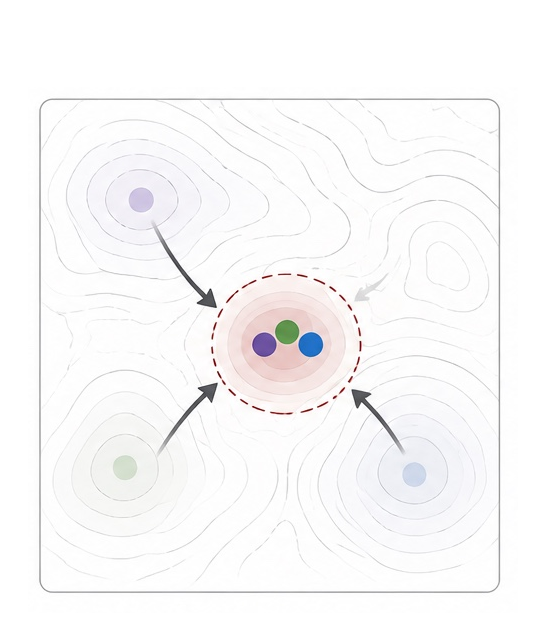}
\caption{Full-solution exchange drives convergence}
\end{subfigure}
\hfill
\begin{subfigure}{0.31\textwidth}
\includegraphics[width=\textwidth]{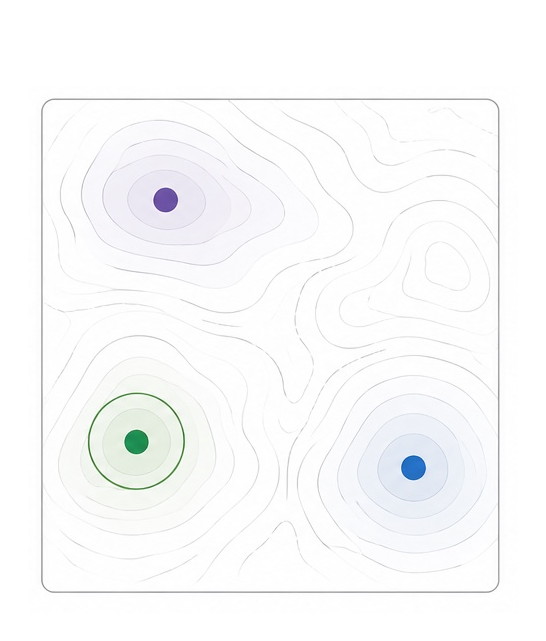}
\caption{Selection preserves diversity}
\end{subfigure}
\caption{\textbf{The interaction tax on optimization tasks.} Different model families (Claude, GPT-4o, Gemini) find structurally different solutions. When agents exchange full candidate solutions, their proposals converge and diversity is lost. When agents generate independently and the best is selected, diversity is preserved.}
\label{fig:concept}
\end{figure*}

\begin{figure*}[t]
\centering
\includegraphics[width=\textwidth]{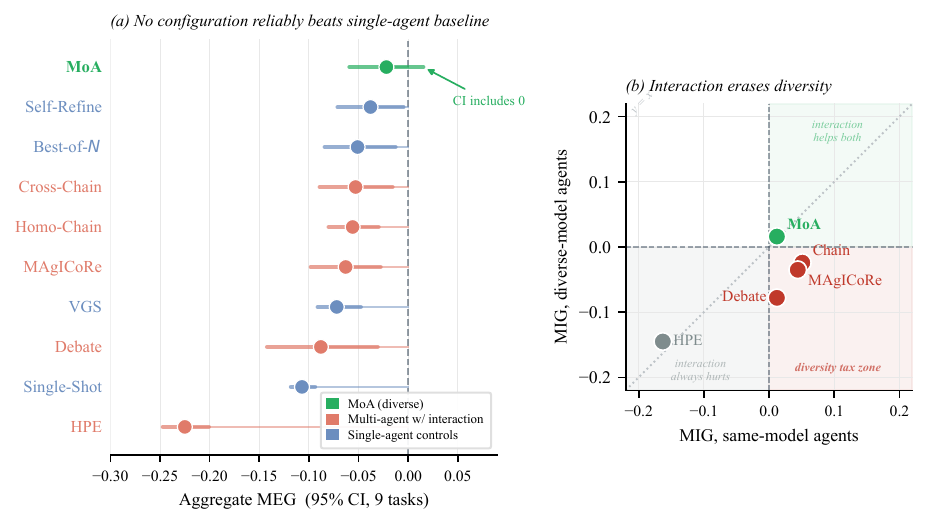}
\caption{(a)~Aggregate MEG for all ten configurations (95\% CI, 9 tasks). MoA is the only configuration whose confidence interval includes zero. (b)~MIG quadrant scatter ($x$-axis = same-model MIG, $y$-axis = diverse-model MIG). MAgICoRe, Chain, and Debate occupy the \emph{interaction tax zone} (bottom-right quadrant), where full-solution interaction helps same-model agents but hurts once models are diverse. MoA escapes to the top-right because its proposers never see each other's outputs.}
\label{fig:meg_mig}
\end{figure*}

\section{Introduction}
\label{sec:intro}

Multi-agent LLM systems are often used because different models may find different solutions. This is especially useful for optimization tasks, where the goal is to find one strong solution from many possible candidates. A diverse team can cover more of the solution space, but only if those differences survive later interaction.

Recent work gives mixed evidence about whether interaction helps. Debate, critique loops, and mixture-of-agents systems report gains from agents reading, reviewing, or synthesizing each other’s outputs \citep{du2024debate,chen2025magicore,wang2025moa}. Other work finds that multi-agent systems can add cost without improving quality under equal budgets \citep{tran2026equal,xu2026single,jarrett2025multiagent}, or that independent sampling already captures much of the benefit of using multiple agents \citep{li2024moreagents}. One reason these results can coexist is that some systems let agents read complete solutions before they answer again, while others keep solutions independent until selection or synthesis.

We study this on 11 verifier-scored optimization tasks. Each agent outputs a candidate solution, and a deterministic verifier returns a scalar score. Claude, GPT-4o, and Gemini each achieve the best score on at least one task, while each same-model team scores zero on at least one task. A diverse set of models can therefore find solutions that a same-model team may miss.

We then test what happens when agents exchange full candidate solutions. When agents read each other’s complete outputs, their later solutions become more similar after one round, as shown in Figure~\ref{fig:concept}. This can help same-model teams, where agents start from similar solutions, but it often hurts diverse-model teams, where the useful difference is erased. We call this loss the \emph{interaction tax}.

The tax does not mean that all communication is harmful. MoA \citep{wang2025moa} avoids the collapse because agents produce candidate solutions independently before synthesis. Critique can also help when it shows a fault that the model can find and repair, as in Knapsack, but is less reliable when the fault is harder to locate, as in 3AP-Free.

Our main contribution is to show that full-solution interaction is a weak default. Idit can erase useful model diversity, while independent generation and repair-focused critique avoid this failure in different ways.

\section{Related Work}
\label{sec:related}

\para{Single vs.\ multi-agent comparisons.}
Multi-agent debate improves factuality \citep{du2024debate}, MAgICoRe raises math accuracy through solver-reviewer-refiner loops \citep{chen2025magicore}, and MoA synthesis dioutperforms individual models on language benchmarks \citep{wang2025moa}. On the other side, Tran and Kiela \citep{tran2026equal}, Xu et al.\ \citep{xu2026single}, Li et al.\ \citep{li2024moreagents}, and Jarrett et al.\ \citep{jarrett2025multiagent} each demonstrate multi-agent losses under various conditions, but all use same-model teams or do not control for model diversity. Our work isolates model diversity as a variable and tests when interaction preserves or erases the solution diversity that diverse models provide.

\para{Evolutionary approaches.}
FunSearch \citep{romeraparedes2024funsearch}, Evolution through Large Models \citep{lehman2023elm}, and AlphaEvolve \citep{georgiev2025alphaevolve} embed LLMs inside evolutionary loops for mathematical discovery. These approaches preserve population diversity through selection pressure rather than through agent interaction, sidestepping the convergence mechanism we study.

\para{Ensemble diversity theory.}
The ensemble diversity decomposition \citep{krogh1994neural,hansen1990neural,lakshminarayanan2017ensembles} shows that ensemble error equals mean member error minus a diversity term. When members communicate and converge, diversity shrinks toward zero. The interaction tax is a structural instance of this effect.

\section{Methods}
\label{sec:formulation}

\para{Setup.} Each task is scored by a deterministic evaluator. Each configuration is a workflow run under identical resource budgets (details in supplementary). Agents see a visible dev evaluator during search. Protocols use this score for refinement, selection, critique, and aggregation. After the run, the dev-selected artifact is evaluated once by a hidden evaluator. Hidden evaluators are stricter, more complete, or robustness-shifted checks designed to test whether a solution found using the visible scorer still holds up offline. We use hidden scores for MEG comparisons and visible feasibility for the two hard-validity analyses. Visible and hidden rankings diverge on three of the tasks analyzed with hidden scores.

We test ten configurations from a larger protocol pool (see supplementary) built from Claude Sonnet~4 \citep{anthropic2025claude}, GPT-4o \citep{openai2024gpt4o}, and Gemini~2.5~Flash \citep{google2025gemini}. Four are single-agent baselines (Single-Shot, Best-of-$N$ sampling, Self-Refine, and Verifier-Guided Search). The remaining configurations are multi-agent workflows spanning sequential revision (Chain), solver-reviewer-refiner critique (MAgICoRe \citep{chen2025magicore}), critique exchange (Debate \citep{du2024debate}), plan-and-execute (HPE \citep{wang2023plansolve}), and independent-proposer synthesis (MoA \citep{wang2025moa}). MoA is the only configuration where proposers never read each other's outputs. The benchmark contains 11 verifier-scored optimization tasks, with four adapted from the AlphaEvolve suite \citep{georgiev2025alphaevolve} (full configuration and task details in supplementary). All tasks share the same interface, where agents output JSON candidate solutions and deterministic verifiers return scalar scores. Tasks differ in feedback structure. On graded-feedback tasks (e.g., MaxCut, Circle Packing), almost every well-formed answer receives a meaningful score. On hard-validity tasks (e.g., Knapsack, 3AP-Free), a structural rule must be satisfied before the score is meaningful, and a small edit can violate the rule and collapse the score. Raw scores are normalized to $Q \in [0,1]$. The main benchmark uses five seeds per cell, and the $2{\times}2$ factorial uses ten seeds on three tasks with sufficient score variance (task subsets noted per figure).

\para{Metrics.} Positive MEG means the configuration outperformed every single-agent baseline. Positive MIG means interaction improved over independent generation. The diverse parallel baseline is inherently stronger because it benefits from coverage, so negative diverse MIG is a conservative finding (formal definitions in supplementary).

\section{Results}
\label{sec:results}

\para{Diversity creates coverage.}
Every same-model team has at least one task with $Q{=}0$ (Table~\ref{tab:coverage}), so no single model covers all tasks. A diverse team (Claude $+$ GPT-4o $+$ Gemini) never scores zero on any task, even though its aggregate Q is comparable to the best same-model team (capability confound check in supplementary). To test whether this coverage translates into a measurable performance advantage, we run a controlled factorial experiment.

\begin{table}[h]
\centering
\caption{Best-of-$N$ Q-scores per model on six tasks where scores diverge. Bold = best model. Each model family wins on different tasks.}
\label{tab:coverage}
\small
\begin{tabular}{@{}lrrrl@{}}
\toprule
Task & Claude & GPT-4o & Gemini & Winner \\
\midrule
Circle Packing   & $\mathbf{1.000}$ & $0.519$ & $0.745$ & Claude \\
TSP-50           & $\mathbf{0.141}$ & ---     & $0.115$ & Claude \\
Erd\H{o}s        & $0.131$          & $\mathbf{0.710}$ & $0.142$ & GPT-4o \\
TSP-100          & $0.013$          & $\mathbf{0.021}$ & $0.010$ & GPT-4o \\
DiffBases        & $0.209$          & $0.264$ & $\mathbf{0.610}$ & Gemini \\
Flat Poly        & $0.000$          & $0.007$ & $\mathbf{0.143}$ & Gemini \\
\bottomrule
\end{tabular}
\end{table}

\para{A controlled check on the diversity advantage.} A $2{\times}2$ factorial crosses model diversity (same vs.\ diverse proposers) with synthesis (MoA synthesis vs.\ best-score selection) on three tasks with sufficient Q-score variance ($N{=}120$, task-stratified bootstrap). The diversity coefficient is $+0.188$ (CI $[+0.073,\;+0.299]$, $p{<}0.001$, factorial table in supplementary). The synthesis coefficient is near zero. Leave-one-out analysis (supplementary) shows the finding is task-dependent (see Limitations).

\para{The interaction tax.} Same-model interaction can refine proposals. Chain, MAgICoRe, and Debate all produce positive same-model MIG (Table~\ref{tab:mig}). With diverse models, all three turn negative (Chain $-0.024$, MAgICoRe $-0.035$, Debate $-0.078$), with bootstrap $P(\text{same}{>}\text{div}) \geq 88\%$. For MAgICoRe and Debate, only the models change across conditions, so the loss is consistent with the diversity erased by full-solution interaction. MoA's MIG remains positive from same-model to diverse ($+0.012 \to +0.016$), which suggests the damage comes from the full-solution interaction step. No configuration achieves positive aggregate MEG (Figure~\ref{fig:meg_mig}).

\begin{table}[h]
\centering
\caption{Mean MIG per configuration family (8 tasks). $P(\text{same}{>}\text{div})$: bootstrap probability that same-model MIG exceeds diverse MIG.}
\label{tab:mig}
\small
\begin{tabular}{@{}lrrr@{}}
\toprule
Configuration & Same MIG & Div MIG & $P$ \\
\midrule
Chain      & $+0.051$       & $-0.024$    & 97\% \\
MAgICoRe   & $+0.044$       & $-0.035$    & 99\% \\
Debate     & $+0.012$       & $-0.078$    & 88\% \\
HPE        & $-0.163$       & $-0.145$    & 38\% \\
MoA        & $+0.012$       & $+0.016$    & 42\% \\
\bottomrule
\end{tabular}
\end{table}

\para{How convergence happens.} This loss is immediate because diversity collapses within a single round of full-solution interaction. On Erd\H{o}s, diverse Debate achieves a strong intermediate score at round~2 but regresses at round~3 once agents read each other's full solutions (round-by-round scores in supplementary). Across protocols, mean pairwise distance between agents' solution representations falls from $0.315$ before interaction to $0.229$ after. The damage happens at the moment of full output exchange, not gradually over multiple rounds. The collapse happens because synthesis does not recombine proposals. On five of seven tasks, synthesis copies the best proposer's output $\geq$80\% of the time rather than combining parts from each (classification table in supplementary). On Erd\H{o}s, GPT-4o outputs the identical trivial constant on 100\% of seeds ($Q{=}0.710$), so any interaction or synthesis step converges to that single point regardless of what Claude or Gemini proposed.

\para{Critique helps only when the fault is easy to locate.} In our experiments, the interaction tax reverses only when the shared information points to a concrete, repairable fault. In Knapsack-50, a failed solution usually violates a capacity constraint, and the critic can propose a direct repair by removing or swapping overweight items. Diverse Debate achieves 10/10 feasibility versus 2/10 for same-model Debate (Table~\ref{tab:constraint_mig}). In contrast, 3AP-Free-100 also has an exact verifier, but locating the violated rule requires finding an arithmetic-progression triple among many combinations. Diverse Debate drops to 0/10 versus 6/10 for same-model. The same qualitative contrast appears under MAgICoRe, though the Knapsack effect is weaker. On graded-feedback tasks, the first critique round degrades solutions 57\% of the time (17/30 runs), consistent with critique offering general guidance rather than a concrete repair (per-run breakdown in supplementary).

\begin{table}[h]
\centering
\caption{Explicit feasibility rates on Knapsack and 3AP-Free ($n{=}10$ per cell). On Knapsack (fault easy to locate), diverse models help. On 3AP-Free (fault hard to locate), diverse models hurt.}
\label{tab:constraint_mig}
\small
\begin{tabular}{@{}llrrl@{}}
\toprule
Task & Config & Same & Diverse & Direction \\
\midrule
\multirow{2}{*}{Knapsack-50}
 & MAgICoRe & 1/10 & 2/10 & $\approx$ \\
 & Debate   & 2/10 & \textbf{10/10} & Div.\ helps \\
\midrule
\multirow{2}{*}{3AP-Free-100}
 & MAgICoRe & 6/10 & 3/10 & Div.\ hurts \\
 & Debate   & 6/10 & \textbf{0/10} & Div.\ hurts \\
\bottomrule
\end{tabular}
\end{table}

\section{Discussion and Conclusion}
\label{sec:discussion}

Our experiments show that full-solution interaction can destroy the diversity it is meant to exploit. When agents read each other's complete outputs, solutions converge within a single round of full-solution interaction. MoA, where proposers never see each other's outputs, is the only configuration whose diverse MIG remains positive.

These results suggest that multi-agent performance depends less on the number of agents than on the information they exchange and when it is exposed. Full candidate solutions create strong convergence. Lower-bandwidth signals, such as scores, method descriptions, or failure causes, may preserve independent exploration while still supporting coordination. Evolutionary approaches like FunSearch \citep{romeraparedes2024funsearch} and AlphaEvolve \citep{georgiev2025alphaevolve} already do this in effect, preserving population diversity through selection pressure rather than through agent interaction. Critique is most useful when it identifies a concrete repair, as in Knapsack, and can be harmful when the relevant error is difficult for the LLM to locate reliably, as in 3AP-Free.

\para{Limitations and future work.} The coverage finding is task-dependent (removing Erd\H{o}s drops the diversity coefficient to $+0.014$ with CI crossing zero). The main benchmark uses only five seeds per condition, and the diversity factorial uses ten seeds on only three tasks. The explicit feasibility analysis rests on two hard-validity tasks, Knapsack and 3AP-Free, although other tasks such as TSP also have hard structural constraints. All tasks are verifier-scored optimization problems, so the interaction tax may not transfer unchanged to open-ended writing, planning, or dialogue settings. Our experiments identify full-solution interaction as a weak default, but they do not directly test which lower-bandwidth information channel is best. A natural next step is a sharing ablation over scores, method descriptions, failure causes, verified summaries, and late crossover. Code and data are available in the supplementary material.

\bibliographystyle{icml2026}
\bibliography{references}

\newpage
\appendix
\renewcommand{\thesection}{\AlphAlph{\value{section}}}
\raggedbottom
\setlength{\textfloatsep}{4pt plus 2pt minus 2pt}
\setlength{\floatsep}{4pt plus 2pt minus 2pt}
\setlength{\intextsep}{4pt plus 2pt minus 2pt}
\setlength{\dbltextfloatsep}{4pt plus 2pt minus 2pt}
\setlength{\dblfloatsep}{4pt plus 2pt minus 2pt}

\section*{Supplementary Material}

This appendix provides workflow details, reproducibility notes, and extended tables and figures for the main paper.

\input{supplementary_body}

\end{document}

%% file: supplementary_body.tex
\section{Workflow Recipe}
\label{app:workflow}

The paper's practical claim is a reusable workflow for AI-assisted math/CS/ML optimization.

\begin{enumerate}
  \item Choose 2--3 diverse model families.
  \item Generate candidate solutions independently with no cross-agent visibility.
  \item Rank candidates with a visible deterministic checker or development evaluator.
  \item If the task exposes local, checkable violations, allow critique-and-revise passes. Otherwise skip interaction and select among the independent candidates.
  \item Verify the chosen artifact with a stricter held-out evaluator or deterministic check.
\end{enumerate}

The reported gains motivating this rule are performance and correctness gains, not a claim of universal time savings. Backbone diversity contributes $+0.188$ in the $2{\times}2$ factorial (though this is driven by the Erd\H{o}s task; see Section~\ref{app:loo}), while critique raises feasibility from 0\% to 47--73\% only on verifiable constraint tasks.

\section{Reproducibility and Verification}
\label{app:repro}

The artifact includes exact prompt templates for all configurations, pinned model identifiers, task instances, deterministic visible and hidden evaluators, saved JSON run traces with raw model outputs, and scripts that reproduce every table and figure from cached results. This makes the workflow inspectable without re-running API calls and re-runnable within a few hours with ordinary API access. Correctness is checked in three layers. Visible evaluators guide search. Hidden evaluators score the final artifact once. The analysis scripts recompute the reported statistics from saved raw outputs. This separation matters because visible and hidden rankings diverge on three of nine optimization tasks. Throughout this appendix, raw task scores follow each task's native direction. Normalized $Q$ is always higher-is-better.

\section{Quickstart Reproduction}
\label{app:quickstart}

The artifact runs on Node.js~$\geq$~20 and Python~$\geq$~3.10. No GPU is required.

\begin{scriptsize}
\begin{verbatim}
# Install
npm install && pip install -r requirements.txt

# Verify saved scores (~3 min, no API key)
npx tsx tools/evaluate.ts \
  --results results/full-v2 \
  --task bench-difference-bases \
  --verifier both

# Paper tables from cached results
python analysis/analyze_bench.py
# -> results/analysis/meg_full.tsv

# Re-run one cell (needs OPENROUTER_API_KEY)
npx tsx src/runner/run-offline.ts \
  --task bench-difference-bases \
  --protocol moa --seeds 1,2,3
\end{verbatim}
\end{scriptsize}

The verification step prints a table of recorded vs.\ recomputed scores; all rows should match exactly. Re-running one task with three seeds takes approximately 5~minutes and costs under \$1 via OpenRouter. Results are written to \texttt{results/offline/}. The full 11-task, 10-protocol benchmark with 10~seeds takes roughly 4~hours. The artifact README contains the complete reproduction guide.

\section{Minimal End-to-End Example}
\label{app:e2e}

Below is a lightly condensed but faithful excerpt from a single MoA run on Molecule QED (seed~1, higher is better). Three proposers generate independently; the synthesizer sees all three outputs and their dev scores.

\smallskip
\begin{tabular}{@{}p{\columnwidth}@{}}
\textbf{Proposer~1} (Claude, devScore\,$= 0.828$) \\
\quad{\small\texttt{\{"smiles": "COc1ccc(Cc2cnc(N)nc2N)cc1"\}}} \\[4pt]
\textbf{Proposer~2} (GPT-4o, devScore\,$= 0.685$) \\
\quad{\small\texttt{\{"smiles": "O=C(C)Oc1ccccc1C(=O)NCc2ccccc2"\}}} \\[4pt]
\textbf{Proposer~3} (Gemini, devScore\,$= 0.547$) \\
\quad{\small\texttt{\{"smiles": "CC(=O)NCCC(=O)O"\}}}
\end{tabular}
\smallskip

\para{Synthesizer (Claude).} Receives all three proposals with scores. Verbatim reasoning:

\begin{quote}
\small
``Agent~1's molecule \texttt{COc1ccc(Cc2cnc(N)nc2N)cc1} contains: a methoxybenzene ring (drug-like aromatic system), a pyrimidine ring with amino groups (common in pharmaceuticals), connected by a simple methylene bridge\ldots\ Agent~2's molecule has lower QED likely due to suboptimal LogP or molecular weight, while Agent~3's simple amino acid derivative lacks the structural complexity needed for high QED. Since Agent~1 already achieved an excellent QED score of 0.8279 with a synthetically accessible molecule that balances all drug-likeness parameters optimally, this represents the superior solution.''
\end{quote}

Final output: \texttt{\{"smiles": "COc1ccc(Cc2cnc(N)nc2N)cc1"\}} (devScore~$= 0.828$, hiddenScore~$= 0.738$).

The synthesizer correctly identified the best proposal using the interpretable molecular structure and large score gap ($0.83 \gg 0.55$). On tasks where one backbone produces a fixed high-scoring constant (e.g., Erd\H{o}s), synthesis always converges to that constant, erasing the diversity of the other proposals (see the Erd\H{o}s transcript in Section~\ref{app:transcript} below).

\section{Artifact Contents}
\label{app:artifact}

The repository at \url{https://github.com/SummerAnn/interaction-tax} contains: (i)~all 11~task instances with embedded dev and hidden evaluators; (ii)~verbatim prompt templates for all 10~configurations (\texttt{src/config/prompts.ts}); (iii)~pinned model identifiers (\texttt{src/config/models.ts}); (iv)~1,556 saved JSON run traces with raw model outputs, dev scores, and hidden scores; (v)~Python analysis scripts that reproduce every table and figure; and (vi)~the offline runner for re-executing any cell from scratch via OpenRouter.

\section{Full Per-Task MEG}
\label{app:pertask}

\begin{table}[H]
\centering
\caption{Per-task MEG for all ten configurations. MaxCut and LJ-$n{=}41$ omitted (all configurations score zero on both tasks).}
\resizebox{\columnwidth}{!}{%
\begin{tabular}{@{}lrrrrrrrrrr@{}}
\toprule
Task & MoA & SR & BoN & XCh & HCh & MAgI & VGS & Deb & SS & HPE \\
\midrule
Circle Pack   & $-0.24$ & $-0.03$ & $0.00$ & $-0.03$ & $0.00$  & $0.00$  & $0.00$ & $-0.52$ & $0.00$  & $-1.00$ \\
DiffBases & $+0.09$ & $-0.11$ & $0.00$ & $-0.11$ & $-0.20$ & $-0.15$ & $-0.13$& $+0.11$ & $-0.20$ & $-0.20$ \\
Flat Poly & $+0.01$ & $-0.06$ & $-0.08$& $-0.08$ & $-0.08$ & $-0.08$ & $0.00$ & $-0.04$ & $-0.08$ & $-0.08$ \\
TSP-100   & $-0.02$ & $-0.02$ & $-0.01$& $-0.01$ & $-0.02$ & $-0.01$ & $0.00$ & $-0.02$ & $-0.01$ & $-0.02$ \\
Erd\H{o}s & $+0.02$ & $0.00$  & $-0.37$& $-0.13$ & $-0.05$ & $-0.15$ & $-0.48$& $-0.33$ & $-0.48$ & $-0.48$ \\
MolQED    & $-0.01$ & $-0.05$ & $0.00$ & $-0.10$ & $-0.12$ & $-0.14$ & $-0.04$& $+0.02$ & $-0.11$ & $-0.24$ \\
TSP-50    & $-0.05$ & $-0.08$ & $0.00$ & $-0.02$ & $-0.04$ & $-0.05$ & $0.00$ & $-0.02$ & $-0.09$ & $-0.02$ \\
\bottomrule
\end{tabular}}
\end{table}

\section{Per-Backbone Coverage}
\label{app:coverage}

\begin{table}[H]
\centering
\caption{Per-backbone Best-of-$N$ Q-scores on six tasks where backbone scores clearly diverge. Bold = best backbone. On Erd\H{o}s, GPT-4o's $Q{=}0.710$ reflects a trivial constant (Section~\ref{app:strategy_class}); Gemini averages $Q{=}0.142$ (4/5 seeds score $Q{=}0$, one seed matches GPT-4o).}
\label{tab:coverage_supp}
\small
\resizebox{\columnwidth}{!}{%
\begin{tabular}{@{}lrrrl@{}}
\toprule
Task & Claude & GPT-4o & Gemini & Winner \\
\midrule
Circle Packing   & $\mathbf{1.000}$ & $0.519$ & $0.745$ & Claude \\
TSP-50           & $\mathbf{0.141}$ & ---     & $0.115$ & Claude \\
Erd\H{o}s        & $0.131$          & $\mathbf{0.710}$ & $0.142$ & GPT-4o \\
TSP-100          & $0.013$          & $\mathbf{0.021}$ & $0.010$ & GPT-4o \\
DiffBases & $0.209$          & $0.264$ & $\mathbf{0.610}$ & Gemini \\
Flat Poly & $0.000$          & $0.007$ & $\mathbf{0.143}$ & Gemini \\
\bottomrule
\end{tabular}}
\end{table}

\section{Capability Confound Check}
\label{app:capability_confound}

\begin{table}[H]
\centering
\caption{MoA-no-synthesis aggregate $Q$-scores across nine optimization tasks ($n{=}10$ per condition). The diverse team and GPT-4o$\times$3 have similar point estimates, but each same-model team catastrophically fails on at least one task ($Q{=}0.000$), while the diverse team never does.}
\label{tab:capability_confound}
\small
\resizebox{\columnwidth}{!}{%
\begin{tabular}{@{}lrl@{}}
\toprule
Configuration & Aggregate $Q$ & Min task $Q$ \\
\midrule
Diverse (Claude $+$ GPT-4o $+$ Gemini) & $0.349$ & $> 0$ \\
GPT-4o$\times$3                         & $0.215$ & $0.000$ \\
Gemini$\times$3                         & $0.250$ & $0.000$ \\
\bottomrule
\end{tabular}}
\end{table}

\section{Critique Regression Rates}
\label{app:regression}

\begin{table}[H]
\centering
\caption{First-round critique regression rate (MAgICoRe, Gemini backbone, $n{=}15$ per task). Constraint = Knapsack-50, 3AP-Free-100. Optimization = Erd\H{o}s, Flat Polynomials.}
\label{tab:regression}
\small
\resizebox{\columnwidth}{!}{%
\begin{tabular}{@{}lrrl@{}}
\toprule
Task type & Regress $0{\to}1$ & Any regress & $n$ \\
\midrule
Constraint (2 tasks) & $\mathbf{0/30}$ (0\%) & 5/30 (17\%) & 30 \\
Optimization (2 tasks) & 17/30 (57\%) & 25/30 (83\%) & 30 \\
\bottomrule
\end{tabular}}
\end{table}

\begin{table}[H]
\centering
\caption{Feasibility success rates on two constraint tasks (Gemini backbone, $n{=}15$ each). Fisher exact $p$ compares MAgICoRe to Best-of-$N$.}
\label{tab:constraint}
\small
\resizebox{\columnwidth}{!}{%
\begin{tabular}{@{}lrrrl@{}}
\toprule
Task            & BoN & Self-Ref & MAgICoRe & $p$ \\
\midrule
Knapsack-50     & 0\%         & 13\%        & 47\%     & $0.003$ \\
3AP-Free-100    & 0\%         & 7\%         & 73\%     & ${<}0.001$ \\
\bottomrule
\end{tabular}}
\end{table}

\section{MIG per Configuration Family}
\label{app:mig}

\begin{table}[H]
\centering
\caption{Mean MIG per configuration family (8 tasks). $P(\text{same}{>}\text{div})$: bootstrap probability that same-model MIG exceeds diverse MIG.}
\label{tab:mig_supp}
\small
\resizebox{\columnwidth}{!}{%
\begin{tabular}{@{}lrrr@{}}
\toprule
Configuration & Same MIG & Div MIG & $P$ \\
\midrule
Chain      & $+0.051$       & $-0.024$    & 97\% \\
MAgICoRe   & $+0.044$       & $-0.035$    & 99\% \\
Debate     & $+0.012$       & $-0.078$    & 88\% \\
HPE        & $-0.163$       & $-0.145$    & 38\% \\
MoA        & $+0.012$       & $+0.016$    & 42\% \\
\bottomrule
\end{tabular}}
\end{table}

\section{$2{\times}2$ Factorial Estimates}
\label{app:factorial}

\begin{table}[H]
\centering
\caption{$2{\times}2$ factorial estimates ($N{=}120$, 10,000 task-stratified bootstrap). Diversity is the only factor whose CI excludes zero.}
\label{tab:2x2}
\small
\resizebox{\columnwidth}{!}{%
\begin{tabular}{@{}lrl@{}}
\toprule
Term                           & Estimate & 95\% CI \\
\midrule
Intercept (same $+$ selection) & $+0.182$ & $[+0.123,\;+0.246]$ \\
Diversity                      & $+0.188$ & $[+0.073,\;+0.299]$ \\
Synthesis                      & $-0.010$ & $[-0.111,\;+0.094]$ \\
Diversity $\times$ Synthesis   & $+0.046$ & $[-0.132,\;+0.226]$ \\
\bottomrule
\end{tabular}}
\end{table}

\section{Leave-One-Out Sensitivity}
\label{app:loo}

\begin{table}[H]
\centering
\caption{Leave-one-out sensitivity of the diversity main effect. Only removing Erd\H{o}s drops the coefficient to near zero.}
\label{tab:loo}
\small
\resizebox{\columnwidth}{!}{%
\begin{tabular}{@{}lrrl@{}}
\toprule
Task withheld & $\hat{\beta}_{\text{div}}$ & 95\% CI & $p$ \\
\midrule
None ($N{=}120$) & $+0.188$ & $[+0.073,\;+0.299]$ & ${<}0.001$ \\
Erd\H{o}s ($n{=}80$) & $+0.014$ & $[-0.109,\;+0.139]$ & $0.84$ \\
DiffBases ($n{=}80$) & $+0.265$ & $[+0.142,\;+0.381]$ & ${<}0.001$ \\
MolQED ($n{=}80$)    & $+0.286$ & $[+0.125,\;+0.442]$ & ${<}0.001$ \\
\bottomrule
\end{tabular}}
\end{table}

\section{Convergence Mechanism}
\label{app:convergence_mechanism}

Step-level trajectories reveal when diversity dies. On Erd\H{o}s, diverse Debate achieves a strong intermediate score at round~2, when agents exchange only critiques, but collapses at round~3 once agents read each other's full solutions. The damage occurs at the moment of full output exchange, not gradually.

\begin{figure}[H]
\centering
\includegraphics[width=\columnwidth]{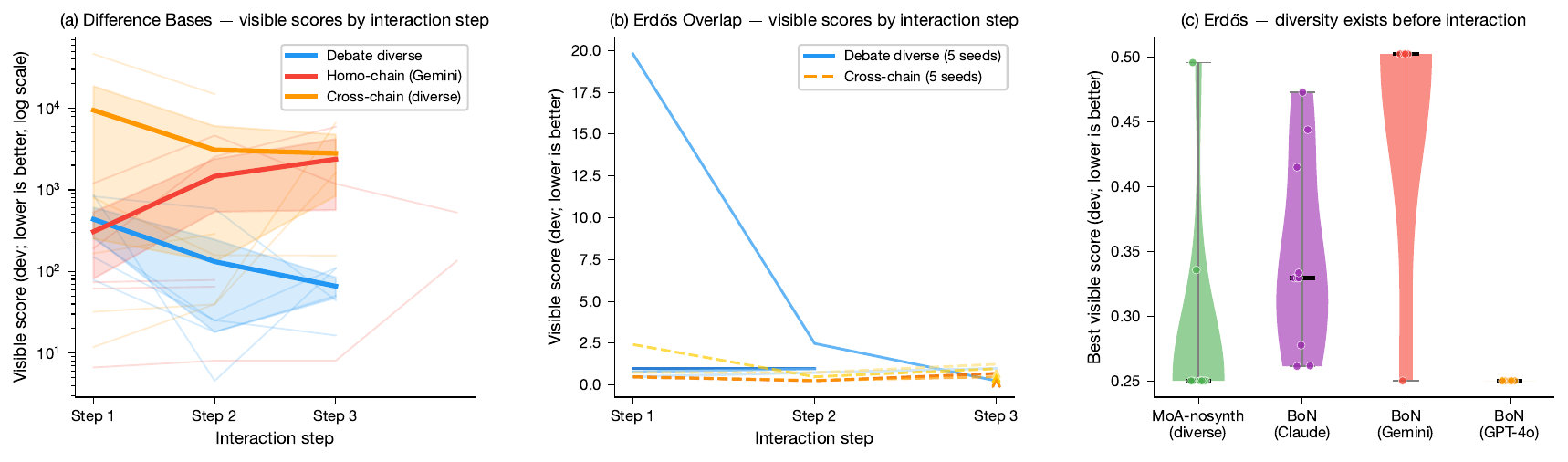}
\caption{Per-round visible-score trajectories on two optimization tasks. (a)~DiffBases: sequential interaction often improves at step~2 but loses that advantage by the final step. (b)~Erd\H{o}s: diverse interaction improves at round~2 then regresses at round~3 after agents exchange full outputs. (c)~Erd\H{o}s: diverse MoA-nosynth achieves broader coverage than same-backbone Best-of-$N$.}
\label{fig:trajectories}
\end{figure}

\section{Diversity Collapse}
\label{app:collapse}

\begin{table}[H]
\centering
\caption{Mean pairwise solution distance. After synthesis, diverse MoA distance ($0.315$) falls to $0.229$.}
\label{tab:diversity}
\small
\resizebox{\columnwidth}{!}{%
\begin{tabular}{@{}lr@{}}
\toprule
Configuration                    & Avg pairwise distance \\
\midrule
MoA (no synthesis)               & $0.315$ \\
Best-of-$N$ (same model)         & $0.312$ \\
Debate                           & $0.309$ \\
MoA (diverse $+$ synthesis)      & $0.229$ \\
\bottomrule
\end{tabular}}
\end{table}

\section{Visible--Hidden Rank Agreement}
\label{app:rank_agreement}

\begin{table}[H]
\centering
\caption{Spearman rank correlation between visible and hidden configuration rankings per task ($n{=}10$ configurations). Three tasks show $\rho < 1$.}
\label{tab:rank_agreement}
\small
\resizebox{\columnwidth}{!}{%
\begin{tabular}{@{}lrlc@{}}
\toprule
Task & $\rho$ & Top-3 overlap & Top-1 inv? \\
\midrule
MaxCut          & $0.891$ & 2/3 & Yes \\
Circle Packing  & $1.000$ & 3/3 & No \\
DiffBases       & $0.806$ & 2/3 & No \\
Flat Poly       & $1.000$ & 3/3 & No \\
TSP-100         & $1.000$ & 3/3 & No \\
LJ-$n{=}41$     & $1.000$ & 3/3 & No \\
Erd\H{o}s       & $1.000$ & 3/3 & No \\
MolQED          & $0.952$ & 3/3 & No \\
TSP-50          & $1.000$ & 3/3 & No \\
\bottomrule
\end{tabular}}
\end{table}

\section{Per-Task MIG Breakdown}
\label{app:pertask_mig}

\begin{table}[H]
\centering
\caption{MIG per task for five configuration families (eight optimization tasks). Bold marks Erd\H{o}s sign flips. On tasks with no Q-score variance (MaxCut, LJ-n41), MIG is zero for all configurations.}
\label{tab:pertask_mig}
\resizebox{\columnwidth}{!}{%
\begin{tabular}{@{}llrrrrrrrr@{}}
\toprule
Family & Cfg & MCut & CPack & DBases & FPoly & TSP1 & LJ41 & Erd\H{o}s & MolQ \\
\midrule
MAgICoRe & homo   & $0.00$ & $0.00$ & $+0.05$ & $0.00$ & $0.00$ & $0.00$ & $\mathbf{+0.33}$ & $-0.03$ \\
MAgICoRe & mixed  & $0.00$ & $0.00$ & $-0.07$ & $0.00$ & $-0.01$ & $0.00$ & $\mathbf{-0.28}$ & $+0.09$ \\
\midrule
Debate & homo     & $0.00$ & $-0.52$ & $+0.31$ & $+0.04$ & $-0.01$ & $0.00$ & $+0.14$ & $+0.13$ \\
Debate & mixed    & $0.00$ & $-0.60$ & $-0.14$ & $+0.02$ & $0.00$ & $0.00$ & $+0.14$ & $-0.04$ \\
\midrule
Chain & homo      & $0.00$ & $0.00$ & $0.00$ & $0.00$ & $-0.01$ & $0.00$ & $\mathbf{+0.43}$ & $-0.01$ \\
XChain & mixed    & $0.00$ & $-0.03$ & $-0.06$ & $-0.03$ & $+0.01$ & $0.00$ & $\mathbf{-0.08}$ & $+0.01$ \\
\midrule
MoA & homo        & $0.00$ & $-0.08$ & $+0.07$ & $0.00$ & $-0.01$ & $0.00$ & $+0.07$ & $+0.04$ \\
MoA & mixed       & $0.00$ & $-0.24$ & $+0.15$ & $+0.06$ & $-0.01$ & $0.00$ & $+0.07$ & $+0.10$ \\
\bottomrule
\end{tabular}}
\end{table}

\section{Agent Configurations}
\label{app:configurations}

\begin{table}[H]
\centering
\caption{Agent configurations. All operate under the same budget vector $(T{=}200\text{K},\,W{=}600\text{s},\,C{=}30\text{s},\,K{=}25)$. HPE = Hierarchical Planner-Executor. Two MoA ablations (same-model proposers, no synthesis) are used in the $2{\times}2$ analysis.}
\label{tab:configurations}
\resizebox{\columnwidth}{!}{%
\begin{tabular}{@{}llll@{}}
\toprule
Configuration & Role & Mechanism & Ref. \\
\midrule
Single-Shot       & Control & One model, one attempt, no iteration. & --- \\
Best-of-$N$ ($N{=}8$) & Control & 8 independent samples, return visible-best. & --- \\
Self-Refine       & Control & Same model generates, critiques, revises for 3 rounds. & Madaan et al., 2023 \\
VGS               & Control & Evolutionary loop (pop.~4, gen.~5, elite~2) with visible scores as fitness. & Romera-Paredes et al., 2024 \\
\midrule
Homo-Chain        & Multi & 4 same-model agents in sequence, each revises prior output. Visible-best returned. & This work \\
Cross-Chain       & Multi & 3-agent chain (Claude $\to$ GPT-4o $\to$ Gemini). Visible-best returned. & This work \\
MAgICoRe          & Multi & Solver $\to$ reviewer $\to$ refiner, 2 rounds. & Chen et al., 2025 \\
Debate            & Multi & 2 agents propose, exchange critiques (2 rounds), synthesizer merges. & Du et al., 2024 \\
HPE               & Multi & Planner decomposes, executors solve subproblems, integrator assembles. & Wang et al., 2023; Hong et al., 2023 \\
MoA               & Multi & 3 proposers generate independently, 1 synthesizer combines. & Wang et al., 2025 \\
\bottomrule
\end{tabular}}
\end{table}

All proposer calls use temperature $0.7$; critique and synthesis calls use $0.3$--$0.5$, consistent across backbones.

\section{Synthesis Decision Quality}
\label{app:synthesis}

\begin{table}[H]
\centering
\caption{Synthesis output classified relative to the best-scoring proposer. On five of seven tasks synthesis copies the best proposal $\geq$80\% of the time. On Difference Bases it degrades 50\% of the time.}
\label{tab:synthesis_decisions}
\small
\resizebox{\columnwidth}{!}{%
\begin{tabular}{@{}lrrr@{}}
\toprule
Task             & Improved & Copied & Degraded \\
\midrule
Circle Packing   & 60\%     & 40\%   & 0\%      \\
Molecule QED     & 20\%     & 80\%   & 0\%      \\
Erd\H{o}s        & 10\%     & 80\%   & 10\%     \\
Flat Polynomials & 0\%      & 100\%  & 0\%      \\
TSP-100          & 0\%      & 100\%  & 0\%      \\
TSP-50           & 0\%      & 90\%   & 10\%     \\
Difference Bases & 7\%      & 43\%   & 50\%     \\
\bottomrule
\end{tabular}}
\end{table}

\section{Within-Backbone Comparisons ($n{=}15$)}
\label{app:within}

\begin{table}[H]
\centering
\caption{Within-backbone comparisons (Gemini, $n{=}15$ per condition). Positive $d$ = iterative configuration is worse.}
\label{tab:within}
\resizebox{\columnwidth}{!}{%
\begin{tabular}{@{}llrrrl@{}}
\toprule
Task & Config & Iter & BoN & $d$ & $p$ \\
\midrule
Flat Poly & MAgI-G & 2.087 & 1.899 & $+1.25$ & $0.002$ \\
Flat Poly & SR-G   & 2.020 & 1.899 & $+1.02$ & $0.009$ \\
TSP-100   & MAgI-G & 53,351 & 50,786 & $+1.42$ & $0.001$ \\
TSP-100   & SR-G   & 53,239 & 50,786 & $+1.46$ & $0.001$ \\
TSP-50    & MAgI-G & 24,167 & 23,103 & $+1.23$ & $0.002$ \\
MolQED    & MAgI-G & 0.863 & 0.859 & $+0.05$ & $0.93$  \\
\bottomrule
\end{tabular}}
\end{table}

\section{Task Descriptions}
\label{app:tasks}

\begin{table}[H]
\centering
\caption{Benchmark tasks. Four optimization tasks are adapted from AlphaEvolve (Georgiev et al., 2025) with smaller parameterizations. Two constraint tasks test interaction under checkable violations.}
\label{tab:tasks}
\resizebox{\columnwidth}{!}{%
\begin{tabular}{@{}lllp{4cm}@{}}
\toprule
Task & Domain & Type & Objective \\
\midrule
Circle Packing & Geometry & Optimization & Maximize radius of $n$ non-overlapping circles in a unit square \\
Difference Bases & Combinatorics & Optimization & Find a set whose pairwise differences cover all residues mod $n$ \\
Erd\H{o}s Overlap & Analysis & Optimization & Minimize $\max_t \int |f(x) f(x{+}t)|\,dx$ over translates \\
Flat Polynomials & Algebra & Optimization & Minimize $\|p\|_\infty$ for a polynomial with $\pm 1$ coefficients \\
TSP-50 / TSP-100 & Routing & Optimization & Minimize tour length for 50 or 100 cities \\
MaxCut & Graph theory & Optimization & Maximize cut value in a fixed graph \\
LJ-$n{=}41$ & Chemistry & Optimization & Minimize Lennard-Jones potential for 41 atoms \\
Molecule QED & Drug design & Optimization & Maximize quantitative estimate of drug-likeness (QED score) \\
Knapsack-50 & Combinatorics & Constraint & Maximize value subject to weight capacity constraint \\
3AP-Free-100 & Combinatorics & Constraint & Find the largest 3-AP-free subset of $\{1,\ldots,100\}$ \\
\bottomrule
\end{tabular}}
\end{table}

MaxCut and LJ-$n{=}41$ are omitted from most analyses because all configurations score $Q{=}0$ on both tasks under the given budget, providing no inter-configuration signal.

\section{Representative Prompt Templates}
\label{app:prompts}

Full prompt templates for all ten configurations are available in the release artifact. Below we show the structure of two representative prompts.

\para{Proposer prompt (used in MoA, Best-of-$N$, Single-Shot).}

\begin{scriptsize}
\begin{verbatim}
System: You are a mathematical optimization
agent. Your goal is to [OBJECTIVE]. Return
ONLY executable Python code that prints the
solution as a JSON object.

User: [Full task description with parameters,
constraints, input data, and evaluation
function specification.]

Output format: a single Python code block.
\end{verbatim}
\end{scriptsize}

\para{Reviewer prompt (used in MAgICoRe, Debate).}

\begin{scriptsize}
\begin{verbatim}
System: You are a mathematical reviewer.
Analyze the proposed solution for correctness.
List specific errors, constraint violations,
or improvements.

User: Task: [OBJECTIVE]
Proposed solution (scored [VISIBLE_SCORE]):
[SOLUTION_CODE]

Provide a numbered list of issues and concrete
suggestions for improvement.
\end{verbatim}
\end{scriptsize}

The proposer prompt is identical across configurations; only the orchestration (whether agents see prior outputs) differs. The reviewer prompt is used only in critique-based configurations. All prompts, including the synthesizer prompt for MoA and the planner/executor prompts for HPE, are included verbatim in the artifact.

\section{Example Multi-Agent Exchange}
\label{app:transcript}

Below is a condensed transcript from a Cross-Chain run on Erd\H{o}s (seed~1). On Erd\H{o}s, \emph{lower} raw scores are better; after Q-normalization, higher~$Q$ is always better.

\para{Example 1. Sequential interaction converges to a degenerate constant (Cross-Chain, Erd\H{o}s, seed~1).}

\para{Step 1 (Claude, dev score $= 0.500$).} Claude reasons about minimizing overlap with translates and produces a structured sparse function using binary patterns.

\para{Step 2 (GPT-4o, dev score $= 0.250$).} GPT-4o sees Claude's solution but outputs a trivial constant $f = 0.5$. GPT-4o defaults to near-constant functions on this task regardless of input context ($Q{=}0.710$, best single backbone on Erd\H{o}s).

\para{Step 3 (Gemini, dev score $= 0.694$).} Gemini produces a structured binary pattern but scores worse than GPT-4o. The chain's visible-best selection picks GPT-4o's output ($0.250$, lowest hence best).

GPT-4o's trivial constant is genuinely the best-scoring solution ($Q{=}0.710$), but it contributes zero strategy diversity because every GPT-4o seed produces the identical output (Table~\ref{tab:strategy_class}). When a chain includes GPT-4o on Erd\H{o}s, the non-trivial solutions from Claude and Gemini are discarded in favor of a degenerate constant. Under independent generation (MoA-nosynth), the same three models produce raw scores of $2.47$ ($Q{=}0$), $0.34$ ($Q{=}0.710$), and $0.50$ ($Q{=}0$); the best ($0.34$, $Q{=}0.710$) is selected, outperforming any single-backbone run.


\section{Per-Task $2{\times}2$ Cell Means}
\label{app:2x2_pertask}

\begin{table}[H]
\centering
\caption{Mean $Q$-scores for each arm of the $2{\times}2$ diversity $\times$ synthesis factorial, broken down by task ($n{=}10$ per cell). The diversity contrast shows that Erd\H{o}s drives the aggregate diversity coefficient, DiffBases contributes moderately, and MolQED contributes near zero. The synthesis contrast is near zero on all three tasks.}
\label{tab:2x2_pertask}
\small
\resizebox{\columnwidth}{!}{%
\begin{tabular}{@{}lrrrrrr@{}}
\toprule
Task & Same$+$Sel & Same$+$Syn & Div$+$Sel & Div$+$Syn & Div contr. & Syn contr. \\
\midrule
DiffBases & $0.202$ & $0.071$ & $0.237$ & $0.287$ & $+0.126$ & $-0.041$ \\
Erd\H{o}s & $0.031$ & $0.071$ & $0.568$ & $0.497$ & $+0.482$ & $-0.016$ \\
MolQED    & $0.314$ & $0.377$ & $0.307$ & $0.437$ & $+0.026$ & $+0.097$ \\
\bottomrule
\end{tabular}}
\end{table}

The per-task breakdown is essential context for the leave-one-out result (Section~\ref{app:loo}). Erd\H{o}s contributes $+0.482$ to the diversity contrast, which is $2.5{\times}$ larger than DiffBases ($+0.126$) and $19{\times}$ larger than MolQED ($+0.026$). Removing Erd\H{o}s therefore drops the aggregate coefficient from $+0.188$ to $+0.014$. The synthesis contrast is near zero or slightly negative on all three tasks, confirming the null synthesis finding is not an artifact of task selection.

\section{Cross-Synthesizer Replication}
\label{app:synthesizer}

\begin{table}[H]
\centering
\caption{Diversity coefficient under three synthesizer backbones. The coefficient remains positive and significant for all three synthesizers. The synthesis coefficient remains near zero regardless of which model synthesizes.}
\small
\resizebox{\columnwidth}{!}{%
\begin{tabular}{@{}lrl@{}}
\toprule
Synthesizer & $\hat{\beta}_{\text{div}}$ & 95\% CI \\
\midrule
Claude Sonnet~4   & $+0.234$ & $[+0.106,\;+0.355]$ \\
GPT-4o            & $+0.176$ & $[+0.049,\;+0.303]$ \\
Gemini~2.5~Flash  & $+0.170$ & $[+0.044,\;+0.298]$ \\
\bottomrule
\end{tabular}}
\end{table}

The diversity coefficient is stable across synthesizer choices. All three synthesizer backbones produce a positive, significant diversity coefficient. This rules out the possibility that the null synthesis effect (Table~\ref{tab:2x2}) is an artifact of a particular synthesizer being too weak or too strong to combine diverse proposals.

\section{Model Strategy Classification}
\label{app:strategy_class}

\begin{table}[H]
\centering
\caption{Strategy frequency per backbone on Erd\H{o}s and DiffBases, classified from rawOutput keywords across all single-shot seeds. Categories are non-exclusive keyword tags. GPT-4o outputs a trivial constant (${\approx}0.005$) on 100\% of Erd\H{o}s runs. Gemini also produces near-trivial constants (all zeros, raw score ${\approx}0.503$). Only Claude generates structurally varied, non-trivial solutions. On DiffBases, GPT-4o uses triangular numbers on every run.}
\label{tab:strategy_class}
\resizebox{\columnwidth}{!}{%
\begin{tabular}{@{}llrrr@{}}
\toprule
Task & Strategy type & Claude & GPT-4o & Gemini \\
\midrule
\multirow{4}{*}{Erd\H{o}s}
 & Trivial constant             & 0/10 & \textbf{5/5} & \textbf{5/5} \\
 & Binary/block pattern        & 4/10 & 0/5 & 0/5 \\
 & Smooth/cosine-based         & 2/10 & 0/5 & 0/5 \\
 & Mathematical reasoning      & 6/10 & 0/5 & 0/5 \\
\midrule
\multirow{4}{*}{DiffBases}
 & Sidon sets ($B_2$ sets)     & 4/10 & 0/5 & 1/5 \\
 & Triangular numbers          & 1/10 & \textbf{5/5} & 1/5 \\
 & Arithmetic progression      & 1/10 & 0/5 & 2/5 \\
 & Greedy/algorithmic          & 2/10 & 0/5 & 0/5 \\
\bottomrule
\end{tabular}}
\end{table}

On Erd\H{o}s, GPT-4o and Gemini each produce a single stereotyped output across all seeds, contributing zero strategy variance to any diverse ensemble. A ``three-backbone'' team effectively has one contributor. On DiffBases, GPT-4o is similarly fixed (triangular numbers on 100\% of runs), though Gemini and Claude each use multiple strategies. The effective diversity of an ensemble depends on whether each backbone explores different regions of the solution space, not merely on the number of distinct model families.

This has a direct consequence for the MIG results. When a diverse Debate or Cross-Chain team includes GPT-4o on Erd\H{o}s, one agent always proposes the same trivial constant ($Q{=}0.710$). Although this constant scores highest, it contributes zero strategy variance, so the nominal three-backbone team has effectively one non-trivial contributor. The interaction step risks discarding Claude's or Gemini's structurally different solutions in favor of this single fixed output, producing the regression visible in the per-task MIG breakdown (Section~\ref{app:pertask_mig}).

\section{Within-Model vs.\ Between-Model Variance}
\label{app:variance_decomp}

\begin{table}[H]
\centering
\caption{Variance decomposition of single-shot $Q$-scores across three backbones ($n{=}5$ seeds each). Between-model variance ($\sigma^2_B$) measures how different the model means are. Within-model variance ($\sigma^2_W$) measures per-seed noise. A $B/W$ ratio above $1$ indicates that models are more different from each other than from themselves, the prerequisite for backbone diversity to help. Two tasks clearly meet this criterion ($B/W \geq 2$) and two are marginal ($B/W \approx 1.2$--$1.3$). LJ-$n{=}41$ and TSP-50 are omitted (LJ because all models score $Q{=}0$, TSP-50 for incomplete backbone coverage).}
\label{tab:variance_decomp}
\resizebox{\columnwidth}{!}{%
\begin{tabular}{@{}lrrrl@{}}
\toprule
Task & $\sigma^2_W$ & $\sigma^2_B$ & $B/W$ & Diversity useful? \\
\midrule
Circle Packing  & $0.017$ & $0.253$ & $15.2$ & Yes \\
MolQED          & $0.015$ & $0.030$ & $2.0$  & Yes \\
Flat Poly       & $0.054$ & $0.070$ & $1.3$  & Marginal \\
Erd\H{o}s       & $0.050$ & $0.061$ & $1.2$  & Marginal \\
TSP-100         & $0.020$ & $0.010$ & $0.5$  & No \\
MaxCut          & $0.0001$& $0.00002$& $0.3$ & No \\
DiffBases       & $0.095$ & $0.016$ & $0.2$  & No \\
\bottomrule
\end{tabular}}
\end{table}

Diversity adds information primarily on tasks where models make structurally different errors ($B/W \geq 2$). On tasks where within-model noise dominates ($B/W < 1$), adding a different backbone contributes no more than resampling the same one. This is consistent with the interaction tax theory. The tax only appears where there is genuine diversity to destroy. Note that $B/W$ measures separation of model \emph{means}, while the $2{\times}2$ diversity contrast measures the gap between the best draw from a diverse team and the best draw from a same-model team. On DiffBases ($B/W{=}0.2$, diversity contrast $+0.126$), within-model noise is high, but Gemini's mean ($Q{=}0.610$) far exceeds Claude's ($Q{=}0.209$), so a diverse team's best draw is still better even though individual seeds are noisy.

\section{Inter-Model Error Correlation}
\label{app:error_corr}

\begin{table}[H]
\centering
\caption{Spearman rank correlations of per-seed hidden scores between model pairs on four tasks with nonzero cross-model Q-score variance. Low or negative $\rho$ indicates uncorrelated errors (diversity helps). Positive $\rho$ indicates correlated errors (diversity adds less). Gemini is constant on Erd\H{o}s (all seeds produce $Q{=}0.000$), precluding correlation.}
\label{tab:error_corr}
\resizebox{\columnwidth}{!}{%
\begin{tabular}{@{}lrrrl@{}}
\toprule
Task & Cl--GPT $\rho$ & Cl--Gem $\rho$ & GPT--Gem $\rho$ & Error indep. \\
\midrule
Erd\H{o}s     & $+0.18$ & --- (const) & --- (const) & Yes \\
DiffBases     & $+0.05$ & $+0.12$ & $+0.17$ & Yes \\
MolQED        & $-0.45$ & $-0.20$ & $-0.33$ & Yes (anti-corr) \\
TSP-100       & $+0.38$ & $+0.50$ & $+0.41$ & No (correlated) \\
\bottomrule
\end{tabular}}
\end{table}

\section{Cross-Backbone Replication}
\label{app:crossbackbone}

The optimization loss pattern replicates on both tested backbones (Gemini and GPT-4o).

\begin{table}[H]
\centering
\caption{MAgICoRe vs.\ Best-of-$N$ on two optimization tasks, two backbones ($n{=}15$ each). The loss replicates on both tested backbones, with large effect sizes.}
\label{tab:crossbb}
\resizebox{\columnwidth}{!}{%
\begin{tabular}{@{}llrrrr@{}}
\toprule
Task & Backbone & MAgICoRe & Best-of-$N$ & $d$ & $p$ \\
\midrule
Flat Poly & Gemini & 2.087 & 1.899 & $+1.25$ & $0.002$ \\
Flat Poly & GPT-4o & 2.951 & 2.251 & $+1.30$ & $0.001$ \\
TSP-100   & Gemini & 53,351 & 50,786 & $+1.42$ & $0.001$ \\
TSP-100   & GPT-4o & 53,529 & 50,128 & $+2.64$ & ${<}0.001$ \\
\bottomrule
\end{tabular}}
\end{table}

\section{Constraint Feasibility Gradient}
\label{app:gradient}

Within the Gemini backbone ($n{=}15$ each), constraint feasibility generally increases with configuration sophistication. Critique-based configurations outperform evolutionary search on both tasks. Single-Shot outperforms Best-of-$N$ on Knapsack (7\% vs.\ 0\%), breaking a strict monotonic gradient at the sampling end.

\begin{table}[H]
\centering
\caption{Constraint feasibility rates by configuration type (Gemini backbone, $n{=}15$). Critique outperforms evolution on both tasks.}
\label{tab:gradient}
\small
\resizebox{\columnwidth}{!}{%
\begin{tabular}{@{}lrrrr@{}}
\toprule
Task & S-Shot & BoN & VGS & MAgICoRe \\
\midrule
Knapsack-50  & 7\%  & 0\%  & 27\% & 47\% \\
3AP-Free-100 & 0\%  & 0\%  & 13\% & 73\% \\
\bottomrule
\end{tabular}}
\end{table}

\section{Constraint MIG: Same-Model vs.\ Diverse Backbones}
\label{app:constraint_mig}

The interaction tax reverses on constraints with trivially checkable violations but persists when verification requires reasoning. On Knapsack-50, where a weight violation is an arithmetic check, diverse Debate achieves 10/10 feasibility versus 2/10 for same-model. On 3AP-Free-100, where verifying a three-term arithmetic progression requires reasoning, diverse Debate drops to 0/10 versus 6/10 for same-model. The pattern replicates across MAgICoRe.

\begin{table}[H]
\centering
\caption{Constraint feasibility rates by backbone condition ($n{=}10$ per cell). On Knapsack (trivially checkable), diverse backbones help. On 3AP-Free (reasoning-required), diverse backbones hurt.}
\label{tab:constraint_mig_supp}
\small
\resizebox{\columnwidth}{!}{%
\begin{tabular}{@{}llrrl@{}}
\toprule
Task & Configuration & Same & Diverse & Direction \\
\midrule
\multirow{2}{*}{Knapsack-50}
 & MAgICoRe & 1/10 & 2/10 & $\approx$ \\
 & Debate   & 2/10 & \textbf{10/10} & Div.\ helps \\
\midrule
\multirow{2}{*}{3AP-Free-100}
 & MAgICoRe & 6/10 & 3/10 & Div.\ hurts \\
 & Debate   & 6/10 & \textbf{0/10} & Div.\ hurts \\
\bottomrule
\end{tabular}}
\end{table}

\section{Formal Metric Definitions}
\label{app:metrics}

Raw scores are normalized to $Q(s) = (s - b)/(r - b)$, clipped to $[0,1]$, where $b$ is the trivial baseline and $r$ the best known result.

Marginal Epistemic Gain (MEG) measures whether a configuration beats the best single-agent baseline:
\[
\meg(p,t) = Q_{\mathrm{hidden}}(p,t) - \max_{c\in\mathcal{C}} Q_{\mathrm{hidden}}(c,t),
\]
where $\mathcal{C} = \{\text{Self-Refine},\text{Best-of-}N,\text{VGS}\}$.

Marginal Interaction Gain (MIG) measures whether interaction helps relative to the same agents running independently:
\[
\mig(p,t) = Q_{\mathrm{hidden}}(p,t) - Q_{\mathrm{hidden}}(\mathrm{parallel}(A),t).
\]

The $2{\times}2$ factorial uses $N{=}120$ runs total (ten per arm on each of the three tasks: Erd\H{o}s, DiffBases, and MolQED).

\section{Step-Level Solution Distance}
\label{app:solution_distance}

\begin{table}[H]
\centering
\caption{Pairwise solution distance between consecutive steps within each run, averaged over seeds. On Debate diverse, distance drops sharply after agents exchange outputs (round~2$\to$3), confirming diversity collapse in real time. Metrics: Jaccard distance for DiffBases (discrete sets), cosine distance for Erd\H{o}s (continuous arrays), Levenshtein distance for MolQED (SMILES strings).}
\label{tab:solution_distance}
\resizebox{\columnwidth}{!}{%
\begin{tabular}{@{}llrrl@{}}
\toprule
Task & Configuration & Step $1{\to}2$ & Step $2{\to}3$ & Pattern \\
\midrule
\multirow{3}{*}{DiffBases} & Debate (same) & $0.950$ & $0.891$ & Slight conv. \\
 & Debate (diverse) & $0.956$ & $\mathbf{0.479}$ & Collapse r3 \\
 & MAgICoRe (div.) & $0.953$ & $0.815$ & Convergence \\
\midrule
\multirow{3}{*}{Erd\H{o}s} & Debate (same) & $0.500$ & --- & Single change \\
 & Cross-Chain & $0.128$ & $0.199$ & Low throughout \\
 & MAgICoRe (div.) & $0.448$ & $0.293$ & Shrinking \\
\midrule
\multirow{2}{*}{MolQED} & Debate (same) & $0.571$ & $0.412$ & Converging \\
 & Debate (diverse) & $0.669$ & $\mathbf{0.378}$ & Fast conv. \\
\bottomrule
\end{tabular}}
\end{table}

\section{Per-Round Debate Scores}
\label{app:debate_rounds}

\begin{table}[H]
\centering
\caption{Per-round visible scores for same-model and diverse Debate on two tasks (lower is better for both, bold = best round). Same-model Debate improves monotonically. Diverse Debate reverses direction at round~3 after agents have exchanged their full outputs.}
\label{tab:traj_scores}
\resizebox{\columnwidth}{!}{%
\begin{tabular}{@{}llrrr@{}}
\toprule
Configuration & Task & Round 1 & Round 2 & Round 3 \\
\midrule
Debate (same)    & DiffBases     & 436  & 132  & \textbf{66}   \\
Debate (diverse) & DiffBases     & 889  & \textbf{72}   & 274  \\
\midrule
Debate (same)    & Erd\H{o}s     & 4.61 & 1.24 & \textbf{0.62} \\
Debate (diverse) & Erd\H{o}s     & 1.28 & \textbf{0.30} & 0.38 \\
\bottomrule
\end{tabular}}
\end{table}

On both tasks, diverse Debate peaks at round~2 (when agents have exchanged critiques but not full solutions) and then regresses at round~3 after full output exchange. Same-model Debate improves monotonically because there is no structural diversity to destroy.

\section{MAgICoRe Step-by-Step Trajectories}
\label{app:magicore_trajectory}

\begin{table}[H]
\centering
\caption{Mean devScore at each MAgICoRe step (step~0 = initial attempt, step~1 = after first critique, step~2 = after second critique), averaged over $n{=}15$ seeds. Direction arrows indicate whether lower ($\downarrow$) or higher ($\uparrow$) is better. $^*$Only 2/15 Erd\H{o}s seeds reach step~2.}
\label{tab:magicore_trajectory}
\resizebox{\columnwidth}{!}{%
\begin{tabular}{@{}lcrrrll@{}}
\toprule
Task & Dir & Step 0 & Step 1 & Step 2 & Trend & Note \\
\midrule
TSP-100   & $\downarrow$ & $54{,}054$ & $53{,}686$ & $53{,}638$ & Slight impr. & Still below BoN \\
Flat Poly & $\downarrow$ & $2.290$ & $2.297$ & $2.345$ & Degrading & Values, not struct. \\
Mol.\ QED & $\uparrow$ & $0.731$ & $\mathbf{0.830}$ & $0.785$ & Peak revert & Step~1 lost \\
Erd\H{o}s  & $\downarrow$ & $0.699$ & $0.504$ & ---$^*$ & Conv.\ trivial & 12/15 $\to$ $0.250$ \\
\bottomrule
\end{tabular}}
\end{table}

On optimization tasks, critique frequently degrades solutions. Molecule QED peaks at step~1 ($+13.5\%$) but reverts at step~2. On Erd\H{o}s, 12 of 15 seeds converge to the trivial constant function at raw score $0.250$ ($Q{=}0.710$). The critic steers Gemini toward the simplest possible answer, which scores highest but is structurally degenerate. Flat Poly shows no structural change across steps. These trajectories are the per-run manifestation of the $0/30$ vs.\ $17/30$ first-round regression rate in Section~\ref{app:regression}.

\section{Proposer Score Spread}
\label{app:proposer_spread}

\begin{table}[H]
\centering
\caption{Mean pairwise score spread among MoA proposers within each run, measuring how different the proposals are before synthesis. Diverse proposers produce $2.2{\times}$ more score spread than same-model proposers on Erd\H{o}s. On DiffBases, within-model variance dominates, so backbone diversity adds less spread.}
\label{tab:proposer_spread}
\small
\resizebox{\columnwidth}{!}{%
\begin{tabular}{@{}lrrl@{}}
\toprule
Task & Diverse spread & Same spread & Ratio \\
\midrule
Erd\H{o}s   & $0.448$ & $0.207$ & $2.2{\times}$ \\
MolQED      & $0.222$ & $0.152$ & $1.5{\times}$ \\
DiffBases   & $1500$  & $3045$  & $0.5{\times}$ \\
\bottomrule
\end{tabular}}
\end{table}

Diversity is real at proposal time. On Erd\H{o}s, diverse proposals are $2.2{\times}$ more spread out than same-model proposals, confirming that different backbones explore different regions of the solution space. On DiffBases, within-model noise dominates ($0.5{\times}$), consistent with the $B/W$ ratio of $0.2$ in Section~\ref{app:variance_decomp}. Synthesis erases this diversity (Section~\ref{app:collapse}).

\section{Relation to Open-Ended Agent Systems}
\label{app:open_ended}

Open-ended platforms such as EinsteinArena \citep{einsteinarena2025} show that multi-agent systems can produce real scientific results. These systems operate in a richer regime than our controlled benchmark, with many attempts over long horizons, public verifiers, leaderboards, discussion, partial results, and accumulated findings. Public descriptions of EinsteinArena and agents such as JSAgent \citep{sung2026einstein} emphasize selective information channels, including scores, notes, failure causes, and reusable findings, rather than only short-horizon peer-to-peer exchange of complete candidate solutions. Our controlled experiments suggest a possible mechanism consistent with such architectures. Selective information may let agents compound useful knowledge without forcing every solver to condition on a complete peer solution early in generation.

\section{Additional Transcript Examples}
\label{app:transcripts_extra}

Examples 2--5 complement Example~1 in Section~\ref{app:transcript}. Together they illustrate synthesis failure, synthesis success, homogenization-as-improvement, and preserved diversity.

\para{Example 2. Synthesis selects a degenerate solution (MoA, Erd\H{o}s, seed~1).}
Three proposers generate independently (lower scores are better on Erd\H{o}s). Claude produces a structured binary-pattern function (devScore~$= 0.490$, $Q{=}0$). GPT-4o outputs a trivial uniform array of constant $0.005$ values (devScore~$= 0.250$, $Q{=}0.710$). The synthesizer selects GPT-4o's output, constructing a mathematical justification:

\begin{quote}
\small
``Agent~2 uses a uniform distribution with constant value $0.005$ across all points. This achieves exactly the theoretical lower bound of $1/4$ for uniform distributions\ldots\ The uniform approach is mathematically optimal\ldots\ Agent~2's solution is superior as it achieves the lower bound with perfect consistency.''
\end{quote}

The synthesizer correctly picked the higher-scoring proposal ($Q{=}0.710$ vs.\ $Q{=}0$), though the mathematical reasoning it gives is wrong. The problem is the loss of strategy diversity. GPT-4o produces this identical trivial constant on every seed (Table~\ref{tab:strategy_class}), so synthesis always converges to the same degenerate output regardless of what Claude or Gemini propose. Diversity existed at proposal time but was erased at synthesis.

\para{Example 3. Synthesis selects the right proposal (MoA, MolQED, seed~1).}
Three proposers produce structurally distinct molecules (higher scores are better on MolQED). Claude produces a methoxybenzene-pyrimidine scaffold (devScore~$= 0.828$). GPT-4o produces an aspirin derivative (devScore~$= 0.685$). Gemini produces a small acetyl amino acid (devScore~$= 0.547$). The synthesizer identifies Claude's molecule as containing ``a methoxybenzene ring (drug-like aromatic system) and a pyrimidine ring with amino groups (common in pharmaceuticals)'' and correctly selects it. Here the quality gap is large ($0.83 \gg 0.55$) and the solution format is chemically interpretable. Synthesis works when the best proposal's advantage is obvious.

\para{Pattern.} On Erd\H{o}s, the synthesizer correctly selects the highest-scoring proposal, but that proposal is a structurally degenerate constant that GPT-4o produces on every seed. On MolQED, the synthesizer also selects the best proposal, but here all three proposals are structurally distinct. The difference is whether diversity survives. When one backbone dominates on score with a fixed output, synthesis erases diversity by always converging to that output. This is consistent with the null synthesis coefficient (Table~\ref{tab:2x2}). Across tasks, synthesis neither reliably helps nor reliably hurts.

\para{Example 4. Homogenization improves score (Debate diverse, DiffBases, seed~1).}
Round~1 (devScore $= 597$). The model reasons creatively about fractal-like structures for a difference basis set. Round~2 (devScore $= 157$). After seeing the first model's approach, the agent shifts to a standard triangular-number construction. Hidden score is $70.2$. This is a positive example of interaction. The solution improved by replacing a creative but poorly executed approach with a conventional mathematical construction. But the improvement came from homogenization. The solution got better \emph{by becoming less creative}. This is exactly the mechanism the diversity coefficient captures at aggregate. Interaction rewards convergence toward standard approaches.

\para{Example 5. Independent generation preserves diversity (MoA-nosynth, Erd\H{o}s, seed~1).}
Three proposers generate independently (lower is better). Claude scores $2.47$ (structured sparse function). GPT-4o scores $0.34$ (compact near-constant solution). Gemini scores $0.50$ (different interpolation). The configuration picks the best (GPT-4o's $0.34$), which is better than any Claude-only attempt on the same seed (best of three is $0.392$). The diverse ensemble finds a solution Claude does not naturally produce. The gain is real, but only because the agents generated without seeing each other.

\section{Protocol Implementation Details}
\label{app:protocol_details}

All ten configurations share the budget vector $(T{=}200\text{K tokens},\,W{=}600\text{s},\,C{=}30\text{s},\,K{=}25)$ and the same task prompt. The selection rule is: the artifact with the best dev score under the task's score direction is submitted for hidden evaluation. Below we summarize implementation details from the release artifact. Full source code is available at \url{https://github.com/SummerAnn/interaction-tax}.

\para{Backbone models.} All models are accessed via OpenRouter: Claude Sonnet~4 (\texttt{anthropic/claude-sonnet-4}), GPT-4o (\texttt{openai/gpt-4o}), Gemini~2.5~Flash (\texttt{google/gemini-2.5-flash}). Model IDs are pinned in the codebase.

\para{Best-of-$N$.} Eight independent calls at temperatures $0.60, 0.65, \ldots, 0.95$. Temperature variation introduces sample diversity without changing the prompt.

\para{VGS.} FunSearch-style evolutionary loop. Initial population of four at varied temperature. Five generations, top-two elites retained, two new candidates per generation via mutation call. Early termination if no improvement for two consecutive generations.

\para{MAgICoRe roles.} Solver generates (SINGLE\_SHOT prompt). Reviewer critiques without producing code (``What are the weaknesses? Be concrete and technical.''). Refiner receives critique and produces improved solution. Two rounds.

\para{Debate.} Debater A uses exploratory strategy (temperature $0.9$): ``Take an EXPLORATORY approach --- try unconventional strategies.'' Debater B uses rigorous strategy (temperature $0.5$): ``Take a RIGOROUS approach --- use well-established methods.'' Two critique rounds, then a neutral synthesizer merges.

\para{MoA.} Three proposers generate independently with no shared context. Aggregator receives all three proposals with dev scores and produces synthesized output. The aggregator prompt includes: ``Multiple agents independently solved this problem. Synthesize their best ideas into a superior solution.''

\para{HPE.} A planner decomposes the problem into sub-foci for parallel executors. Each executor works only on its assigned focus and produces a partial contribution (free-form text, not a complete solution). An integrator assembles the contributions into one schema-valid solution. Multiple rounds give the planner score feedback.

\para{Retry behavior.} All protocols share a common retry mechanism. If the LLM produces malformed JSON or a verifier-rejected solution, the error message is appended to the next call. Generation steps allow up to 5 retries; refinement steps allow up to 2 to conserve budget. A cell that exhausts retries at the first generation step is excluded from analysis.

\para{Budget enforcement.} A shared BudgetEnforcer tracks cumulative token usage, wall clock time, and evaluator call count. Any call exceeding a cap is silently dropped and the protocol terminates early.

\section{Extended Prompt Templates}
\label{app:prompts_extended}

Section~\ref{app:prompts} shows two representative prompts. Here we include additional templates verbatim from the release artifact.

\para{MoA aggregator prompt.}

\begin{scriptsize}
\begin{verbatim}
System: You are an Aggregator. Multiple agents
independently solved this problem. Synthesize their
best ideas into a superior solution. Return ONLY
valid JSON.

User: # Problem
[PROBLEM]

# Schema
[SCHEMA]

# Agent Solutions
[CANDIDATES with dev scores]

Synthesize the best solution. Output ONLY JSON.
\end{verbatim}
\end{scriptsize}

\para{Debate synthesizer prompt.}

\begin{scriptsize}
\begin{verbatim}
System: You are a neutral Synthesizer. Two agents
debated this problem. Combine the best ideas from
both into an optimal solution. Return ONLY valid
JSON.

User: # Problem
[PROBLEM]

# Schema
[SCHEMA]

# Debater A (score: [SCORE_A])
[POSITION_A]

# Debater B (score: [SCORE_B])
[POSITION_B]

Synthesize the best solution. Output ONLY JSON.
\end{verbatim}
\end{scriptsize}

\para{MAgICoRe reviewer prompt.}

\begin{scriptsize}
\begin{verbatim}
System: You are a REVIEWER. Analyze the solution
below and provide specific, actionable critique.
What are the weaknesses? What strategies could
improve the score? Be concrete and technical.

User: # Problem
[PROBLEM]

# Current Solution (score: [SCORE], goal: [DIR])
[SOLUTION]

Provide your critique.
\end{verbatim}
\end{scriptsize}

\para{HPE planner prompt.}

\begin{scriptsize}
\begin{verbatim}
System: You are a PLANNER. Decompose this problem
into focused sub-tasks for [N] executor agents
working in parallel. Each executor works ONLY on
its assigned sub-focus, then a separate integrator
will combine the contributions into one final
solution.

Return ONLY valid JSON:
{"strategy": "one-sentence plan",
 "subtasks": [{"id": 1, "focus": "..."},
              {"id": 2, "focus": "..."}, ...]}

Make the foci genuinely complementary -- they should
NOT all attempt the whole problem.
\end{verbatim}
\end{scriptsize}

All ten protocol prompt sets (including VGS mutation, Self-Refine iteration, Homo-Chain refinement, Cross-Chain refinement, HPE executor, and HPE integrator) are committed verbatim in the artifact at \texttt{src/config/prompts.ts}.